# OPTICAL STOCHASTIC COOLING EXPERIMENT AT THE FERMILAB IOTA RING*

J. D. Jarvis[†], V. Lebedev, H. Piekarz, A. L. Romanov, J. Ruan,
Fermi National Accelerator Laboratory, Batavia, IL 60510-5011, USA
M. B. Andorf, P. Piot[1], Northern Illinois University, Dekalb, IL 60115, USA
[1]also at Fermi National Accelerator Laboratory, Batavia, IL 60510-5011, USA

*Abstract*

Beam cooling enables an increase of peak and average luminosities and significantly expands the discovery potential of colliders; therefore, it is an indispensable component of any modern design. Optical Stochastic Cooling (OSC) is a high-bandwidth, beam-cooling technique that will advance the present state-of-the-art, stochastic cooling rate by more than three orders of magnitude. It is an enabling technology for next-generation, discovery-science machines at the energy and intensity frontiers including hadron and electron-ion colliders. This paper presents the status of our experimental effort to demonstrate OSC at the Integrable Optics Test Accelerator (IOTA) ring, a testbed for advanced beam-physics concepts and technologies that is currently being commissioned at Fermilab. Our recent efforts are centered on the development of an integrated design that is prepared for final engineering and fabrication. The paper also presents a comparison of theoretical calculations and numerical simulations of the pickup-undulator radiation and its interaction with electrons in the kicker-undulator.

## INTRODUCTION

Beam cooling compresses a beam's phase space by damping incoherent particle motions. It is a principal means of increasing achievable luminosity, preventing emittance growth due to intra-beam scattering (IBS) and other effects, reducing beam losses and improving energy resolution; therefore, it is an indispensable component of any modern collider design. Beam cooling is an expansive area of research with many notable subfields, e.g. radiation, ionization, electron and stochastic cooling.

Van der Meer's Nobel-winning Stochastic Cooling (SC) was vital in the accumulation of antiprotons and in the delivery of the beam quality required for the discovery of the W and Z bosons [1,2]. In SC and its variants, signals from electromagnetic pickups, operating in the microwave regime with a bandwidth on the order of several GHz, are used in negative feedback systems to reduce the phase-space volume of a circulating beam in all degrees of freedom [1-6].

If every beam particle's deviation from the reference particle could be sensed and corrected individually, then the total error in the beam could be removed in a single pass through a SC system. In practice, the spectral bandwidth, $W$, of the feedback system (pickup, amplifier, kicker) sets a Fourier-limited temporal response $T \sim 1/2W$, which is very large compared to the intra-particle spacing and limits the achievable cooling rate. With a limited bandwidth on the order of several GHz, conventional SC systems become ineffective for the high-density beams of modern colliders. The realization of high-bandwidth/fast cooling techniques, and their translation into operational systems, is a technological imperative for many future colliders.

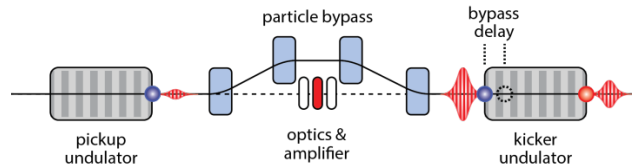

Figure 1: Simplified conceptual schematic of an optical stochastic cooling section. A wavepacket produced in the pickup subsequently passes through transport optics and an optical amplifier. In the kicker undulator, each particle receives an energy kick proportional to its momentum deviation.

## OPTICAL STOCHASTIC COOLING

One possible solution is the extension of the SC principle to optical frequencies ($\sim 10^{14}$ Hz). This would increase cooling rates by three to four orders of magnitude, and would be an extraordinary advance in beam-cooling technology. OSC was first suggested in the early 1990s by Zolotorev, Zholents and Mikhailichenko, and replaced the microwave hardware of SC with optical analogs, such as wigglers and optical amplifiers [7,8]. A number of variations on the original OSC concept have been proposed, and its use has been suggested for hadron, heavy-ion, electron-ion and muon colliders and also controlling emittance growth in electron storage rings [9-15]. At present, a proof-of-principle demonstration with protons or heavy ions involves prohibitive costs, risks and technological challenges [16]; however, demonstration of OSC with medium-energy electrons is a cost-effective alternative that enables detailed study of the beam-cooling physics, optical systems and diagnostics [17-20].

In the transit-time method of OSC, shown schematically in Fig. 1 and upon which this program is based, a particle's deviations from the reference particle are encoded in its arrival time at the kicker system by a magnetic bypass [8]. The particle (an electron for purposes of discussion) first emits a radiation packet while traversing a pickup undulator (PU), the wavelength of which is given by the usual (planar) undulator relation

$$\lambda_r = \frac{\lambda_u}{2n\gamma^2}\left(1 + \frac{K^2}{2} + \gamma^2\theta^2\right), \quad (1)$$

where $\lambda_u$ is the undulator period, $n$ is the harmonic number, $K$ is the undulator-strength parameter and $\theta$ is the observation angle relative to the beam axis.

The radiation packet is transported with (active) or without (passive) amplification to a kicker undulator where it interacts with the same electron. Between the pickup and kicker, the electron traverses a bypass (chicane), which is designed such that a reference particle at the design energy will arrive at the kicker undulator simultaneously with the head of its radiation packet. The energy of the reference particle is unchanged by its interaction with the radiation field in the kicker; however, in the linear approximation, all other particles will have a delay that is proportional to their momentum deviation, $\Delta p/p$, and will receive corresponding corrective kicks towards the design energy.

In this arrangement, the magnitude of the momentum kick received by the particle in a single pass through the cooling system is approximately

$$\frac{\delta p}{p} = -\kappa \sin(k_0 s) \quad (2)$$

where $k_0 = 2\pi/\lambda_r$ and $s$ is the longitudinal displacement of the particle relative to the reference particle after traversing the bypass [16]. The gain parameter $\kappa$ is given by $\kappa = \sqrt{G}\,\Delta E/cp$, where $G$ is the power gain of the optical amplifier and $\Delta E$ is the maximum energy exchange for a particle in the kicker. The form of $\Delta E$ can be estimated analytically by expressing the electric field via the Liénerd-Wiechert relation, and a modified Kirchhoff formula, and subsequently integrating against the motion of a particle phased for maximum kick. This procedure yields

$$\Delta E \approx \frac{\pi q_e^2 N_u}{3\epsilon_0 \lambda_r}K^2\left(1 + \frac{K^2}{2}\right)f_T(K,\gamma\theta_m) \quad (3)$$

as the maximum energy exchange in the kicker without optical amplification. In equation (6), $N_u$ is the number of periods per undulator, and $0 < f_T < 1$ is a correction factor that accounts for longitudinal motions due to large values of $K$ (>1) in the kicker and the finite collection angle of the light optics, $\theta_m$. It is interesting to note that for small $K$ and $f_T = 1$, Equation (6) is identical to twice the energy radiated, in the fundamental band, in a single undulator [21]. This is the amount of radiation detected downstream if the optical delay is detuned by more than the duration of the wavepacket; sweeping the delay through the wavepacket will produce coherent oscillations in the detected radiation of amplitude $\sim \sqrt{G}\,\Delta E$ [22,23]. This is an important diagnostic for temporal synchronization of the light and particle optics.

*Cooling Rates and Ranges*

The most critical parameters in OSC physics are the cooling rates and ranges for the longitudinal and transverse phase planes. In the following analysis we use the usual generalized curvilinear definitions for coordinates and derivatives, and we closely follow the analysis of reference [16]. In the interest of brevity, we enumerate only the salient steps and results. In the linear approximation, a given electron is displaced longitudinally from the reference particle by

$$s = M_{51}x + M_{52}x' + \left(M_{56} - \frac{L_{pk}}{\gamma^2}\right)\frac{\Delta p}{p}, \quad (4)$$

where $(x, x')$ is the transverse position and angle of the particle, $L_{pk}$ is the pickup to kicker distance and $M_{51}$, $M_{52}$, $M_{56}$ are the elements of the linear transfer matrix from the exit of the pickup to the entrance of the kicker. The transverse coordinates can be decomposed as betatron and dispersive components as $x = x_\beta + D\Delta p/p$ and $x' = x_\beta' + D'\Delta p/p$, where $D$ and $D'$ are the dispersion and its derivative along the design orbit at the exit of the pickup. Neglecting betatron oscillations, we have

$$s = \left(M_{51}D + M_{52}D' + M_{56} - \frac{L_{pk}}{\gamma^2}\right)\frac{\Delta p}{p} = S_{pk}\frac{\Delta p}{p}, \quad (5)$$

and so for small $s$, the approximate cooling rate for the longitudinal emittance is then

$$\lambda_s = f_0 \kappa k_0 S_{pk}, \quad (6)$$

where $f_0$ is the revolution frequency in the ring. Redistribution of the cooling between phase planes does not change the sum of cooling rates [16,24]; it can be shown that the horizontal cooling rate is then

$$\lambda_x = f_0 \kappa k_0 \left(M_{56} - S_{pk} - \frac{L_{pk}}{\gamma^2}\right). \quad (7)$$

Neglecting $L_{pk}/\gamma^2$, which will be small in our case (<10$^{-4}$), we see that the ratio of horizontal and longitudinal cooling rates for a given delay in the chicane is determined entirely by the dispersion and its derivative at the exit of the pickup undulator as

$$\frac{\lambda_x}{\lambda_s} = \frac{M_{56}}{S_{pk}} - 1. \quad (8)$$

Coupling between the $x$ and $y$ dimensions can be provided in the ring, outside of the cooling section, thus providing cooling in all degrees of freedom.

In the case of a particle undergoing betatron oscillations, we can parameterize the argument of equation (5) in terms of normalized betatron and synchrotron amplitudes and phases, $(a_x, \psi_x)$ and $(a_p, \psi_p)$ respectively, as

$$\frac{\delta p}{p} = -\kappa \sin(a_x \sin\psi_x + a_p \sin\psi_p). \quad (9)$$

From equations (5) and (8), we find that the normalized synchrotron amplitude is

$$a_p = -k_0(M_{51}D + M_{52}D' + M_{56})\left(\frac{\Delta p}{p}\right)_m, \quad (10)$$

where $(\Delta p/p)_m$ is the amplitude of the synchrotron oscillations. Writing the particle coordinates in the action-angle form and invoking the Courant-Snyder invariant, $\tilde{\varepsilon}$, and the other Twiss parameters ($\beta, \gamma, \alpha$) we arrive at

$$a_x = -k_0\sqrt{\tilde{\varepsilon}(\beta M_{51}^2 - 2\alpha M_{51}M_{52} + (1+\alpha^2)M_{52}^2/\beta)}, \quad (11)$$

By averaging the kick over betatron and synchrotron oscillations (over $\psi_x$ and $\psi_p$), we can determine an average reduction in cooling force. It is given by

$$\begin{bmatrix} F_x \\ F_s \end{bmatrix} = \begin{bmatrix} \lambda_x(a_x,a_p)/\lambda_x \\ \lambda_s(a_x,a_p)/\lambda_s \end{bmatrix} = 2\cos(k_0 s_0)\begin{bmatrix} J_0(a_p)J_1(a_x)/a_x \\ J_0(a_x)J_1(a_p)/a_p \end{bmatrix}, \quad (12)$$

where $J_0$ and $J_1$ are Bessel functions of the first kind, and the $\cos(k_0 s_0)$ term provides for a timing offset between the reference particle ($a_x, a_p = 0$) and its radiation [16,23]. For the reference particle, the bypass will be tuned such that $k_0 s_0 = 0$; however, changes of this delay will produce coherent oscillations in the observed undulator-radiation power, allowing us to determine the optimum setting for cooling. Fig. 2 illustrates the implications of equation (15) for the phase-space dynamics of the particles. The zeros of the Bessel functions comprise various fixed points that establish the cooling ranges of the OSC process. In effect, this normalized ($a_x, a_p$) space has many cooling zones and stable points to which the particles will be driven. In the ideal case, all beam particles should lie within the cooling zone established by the first zero of $J_0$, $\mu_1 = 2.405$, and will be cooled in both planes until reaching equilibrium with quantum excitation by SR. In this case, the longitudinal and transverse cooling ranges, $n_{\sigma p}$ and $n_{\sigma x}$, are determined by simply solving equations (13) and (14) for $(\Delta p/p)_m$ and $\sqrt{\tilde{\varepsilon}}$, with $(a_x, a_p) = \mu_1$, and then normalizing to the beam's rms values, $\sigma_p$ and $\sqrt{\varepsilon_x}$.

We note that when nonlinear path lengthening is included in Equation (7), this cooling surface will be modified for large betatron and synchrotron amplitudes. Note also that the strength and polarity of the heating and cooling zones in Fig. 3 can be modified by changing $k_0 s_0$ [23], and tomographic methods may, in principle, be used to observe the interesting phase space structure that results from equation (15) [25,26].

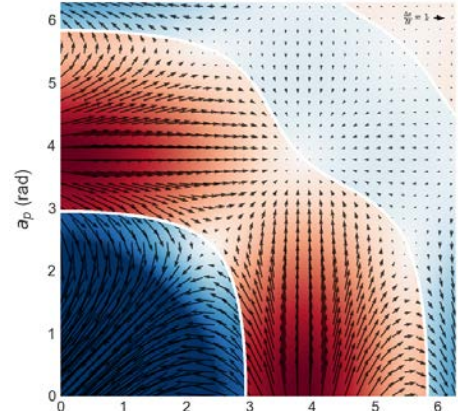

Figure 2: Dependence of the net cooling force on normalized betatron and synchrotron amplitudes in the linear approximation for $k_0 s_0 = 0$. Vector field shows direction of the net cooling/heating.

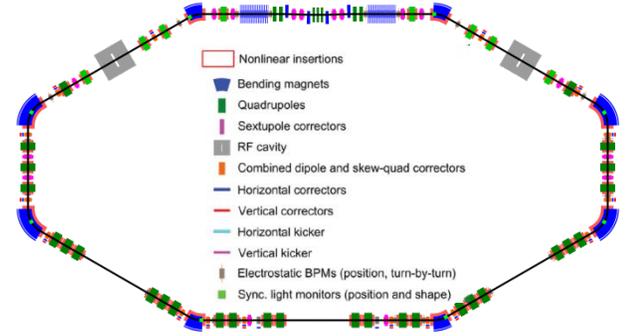

Figure 3: Schematic diagram of the Integrable Optics Test Accelerator ring at FNAL. The base configuration shown here is for the demonstration of nonlinear integrable optics; in OSC configuration, the lower straight is replaced with the insertion shown in Fig. 4.

## OSC AT THE IOTA RING

The Integrable Optics Test Accelerator (IOTA) ring, currently being commissioned at the Fermi National Accelerator Laboratory (FNAL), is a unique test facility for advanced beam-physics concepts and technologies [27]. IOTA's scientific program targets fundamental advancements in beam optics, beam cooling and space-charge compensation, and a robust capability to demonstrate OSC physics was used as a key requirement in the ring's design. The ring is shown schematically in its OSC configuration in Fig. 3.

The IOTA OSC demonstration is unique in that the OSC-damping rate will dominate the ring's synchrotron-radiation damping by a factor of ~60 in the absence of any optical amplification. This means that fundamental OSC physics can be thoroughly explored, early in the experimental program and decoupled from any amplifier development and integration. This is a major advantage relative to other machines where a demonstration of OSC with electrons has been proposed [17-20].

## Electron Optics

The IOTA OSC experiment, shown schematically in Fig. 4, will occupy the straight section at the top of Fig 3. The four dipoles in the electron bypass (B1-B4) have parallel edges to eliminate geometric focusing; in this case $M_{56} = S_{pk}$, and by equation (11), the coupling of the longitudinal and transverse OSC rates vanishes [16]. To introduce p-x coupling we place a defocusing quadrupole (QX) of strength $\Phi = 1/F$ in the center of the bypass. In this configuration, the cooling ratio is approximately $\lambda_x/\lambda_s \approx \Phi D^* h / (2\Delta s - \Phi D^* h)$, and for $\lambda_x/\lambda_s = 1$ the cooling ranges (corresponding to $(a_x, a_p) = \mu_1$) are approximately

$$n_{\sigma s} \approx \frac{\mu_1}{k_0 \sigma_p \Delta s} \qquad n_{\sigma x} \approx \frac{\mu_1}{2k_0 \Delta s} \sqrt{\frac{\mathcal{H}^*}{\varepsilon_x}}, \qquad (13)$$

where $\Delta s$ is the optical delay for the focusing and amplification systems, $h$ is the horizontal trajectory offset in the chicane, $D^*$ and $\mathcal{H}^* = D^{*2}/\beta^*$ are the dispersion and dispersion invariant (assuming vanishing derivatives for the dispersion and beta-functions in the chicane center), and an asterisk denotes that the value is taken in the center of the chicane. Note that for equal cooling rates, $\Phi D^* h = \Delta s$. While operating with shorter wavelength and longer delay increases the cooling rate, it will result in a corresponding reduction in the effective cooling range.

As discussed in [16], the equations in (16) suggest that for a fixed wavelength and delay, we should optimize the optical lattice to maximize the dispersion invariant in the cooling section and minimize the emittance. The equilibrium emittance grows with the average dispersion invariant in the ring due to quantum excitation, so we should increase the invariant in the cooling section and then reduce it as quickly as possible. Note that our low design energy, 100 MeV, also reduces the equilibrium emittance due to quantum excitation and provides correspondingly larger cooling range. For reasonable levels of dispersion, we should then minimize $\beta^*$ using "collider-style" optics [16]; the quadrupoles (Q1-Q4) on either side of the bypass provide a small waist in the chicane center and a negative-identity mapping between the pickup (U1) and kicker (U2) in the x-plane. Finally, sextupole pairs in the bypass provide chromaticity correction to minimize second-order path lengthening, which would otherwise reduce the achievable cooling ranges [28].

## Light Optics

As shown in Fig. 3, there are two configurations for the passive optical system: three-lens and single lens telescopes. Initial designs considered the use of telescopic optics to suppress depth-of-field effects arising from the finite length of the undulators; however, it was discovered that due to slight overfocusing and reduced chromaticity, the maximum achievable kick for a single-lens telescope was nearly identical to that of the three-lens telescope. The single-lens configuration greatly reduces the complexity of the optical system's design, engineering and operation, and provides for an accelerated timeline with significantly reduced risk. For the active configuration, the radiation must be tightly focused in the center of the telescope for efficient amplification, and the simplest two-lens telescope results in a transfer matrix close to positive identity. Experiments at IOTA will use the single-lens and two-lens telescopes for passive and active cooling, respectively.

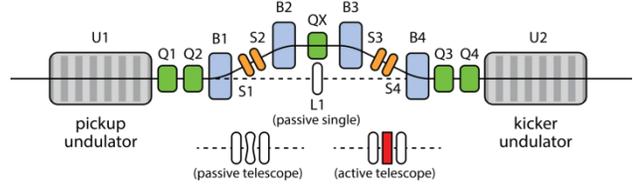

Figure 4: Conceptual schematic of the IOTA OSC insertion with various configurations for the light optics.

Table 1: Design Parameters and Performance Estimates for 0.95-µm and 2.20-µm OSC Configurations in IOTA

| Design wavelength, $\lambda_r$ | 0.95 µm | 2.20 µm |
|---|---|---|
| Number of particles | $10^3$-$10^6$ | |
| Beam kinetic energy | 100 MeV | |
| Delay in the chicane, $\Delta s$ | 0.648 mm | 2.00 mm |
| Offset in the chicane, $h$ | 20.0 mm | 35.1 mm |
| Momentum spread (rms), $\sigma_p$ | 1.00x$10^{-4}$ | 1.06x$10^{-4}$ |
| uncoupled x-emittance (rms); no OSC, $e$ | 1.02 nm | 2.62 nm |
| Beta function in chicane center, $\beta^*$ | 0.25 m | 0.12 m |
| Disp. in chicane center, $D^*$ | 0.27 m | 0.48 m |
| Disp. invariant in chicane center, $A^*$ | 0.29 m | 1.92 m |
| Undulator period, $\lambda_u$ | 47.77 cm | 110.6 cm |
| Number of und. periods, $N_u$ | 7 | 16 |
| On-axis undulator field, $B_0$ | 2.327 kG | 1.005 kG |
| Maximum energy kick, $\Delta E$ | 91.1 meV | 19.6 meV |
| Cooling rates ($\lambda_x$, $\lambda_s$) | (66, 64) s$^{-1}$ | (22, 19) s$^{-1}$ |
| Cooling ranges ($\lambda_x$, $\lambda_s$) | (5.61, 4.73) | (3.97, 5.7) |
| Sync. rad. Damping rates ($x$, $s$) | (0.5, 1.02) s$^{-1}$ | (0.53, 0.91) s$^{-1}$ |

## Configurations

We have designed OSC systems for two distinct operating wavelengths at IOTA: 0.95 µm and 2.2 µm. The general parameters for passive cooling in these configurations are given in Table 1. The estimated energy kicks include the effects of dispersion and depth of field. Operation at 0.95 µm requires a reduction in delay to preserve the cooling range. The reduced delay does not support the use of a multi-lens telescope; therefore, only the passive, single-lens configuration is possible in this case. While the achievable cooling rate is significantly lower in the 2.2-µm case, there is sufficient delay available to implement a simplified single-stage optical amplifier; however, the performance of amplifiers in the mid IR is significantly lower than those in the visible. We estimate that a single pass amplifier based on amplified spontaneous emission in Cr:ZnSe will only increase the cooling force by a factor of 1.65 [29].

Our initial experimental efforts will focus on the 0.95 µm passive configuration due to its higher cooling rate (~2 times greater than the amplified configuration at 2.2 µm), the superiority of optical detectors at this wavelength and enhanced compatibility with other planned studies in IOTA, which involve the storage and characterization of a single electron. Additionally, Due to the larger beta function in the bypass, the transverse angles of the particles are smaller and non-linear path lengthening is correspondingly reduced. This makes sextupole correction less critical than in the 2.2-μm case and enables the use of shorted versions of the existing IOTA sextupoles. Fig. 5 presents an example Synchrotron Radiation Workshop (SRW) simulation of the electric field experienced in the kicker by an electron phased for maximum energy exchange; the corresponding electron trajectory is shown as well. The estimated energy kicks compare well between our SRW simulations and analytic theory. For example, in our 2.2-μm passive configuration the maximum achievable energy exchange, in the absence of dispersion, is 22.0 meV and 20.1 meV in the theory and simulations respectively. A detailed comparison of our SRW simulations with analytic theory is given elsewhere [29].

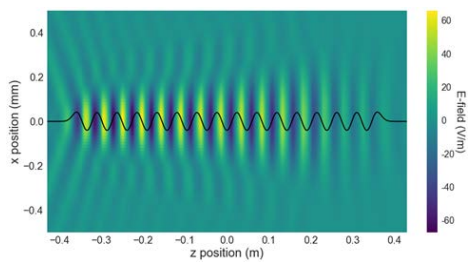

Figure 5: 2D map of the electric field experienced by an electron phased for maximum energy exchange

*Hardware*

The OSC experiment requires the design, engineering and construction of a variety of new hardware, including magnets (dipoles, quadrupoles, sextupoles and correctors) and undulators, light optics and support systems and specialized vacuum chambers and beam pipes. We briefly describe some aspects of each and their current status.

**Magnets:** The dipole magnets require high integrated field quality (~$10^{-4}$) over an aperture of ~5-mm in radius. This can be achieved by use of a monolithic core that is electro-discharge machined to 10-μm precision. An engineering design of the dipole has already been produced and is shown in Fig. 6. The sextupoles for the 0.95-μm experiment are modified versions of the IOTA sextupoles, also shown in Fig. 6, and minimal additional design work will be required. The coupling quad in the center of the chicane will be a Panofsky type and will double as a vertical corrector. Screens are required on most magnetic elements due to their close packing in the bypass. Preliminary magnetic designs for the undulators have been developed and are being optimized to account for saturation effects and thermal considerations.

**Light Optics:** The tolerances on the light optics are relatively relaxed compared to what is available from manufacturers (central thickness, radius of curvature, etc…). For example, in the 2.2-μm case using a three-lens telescope and $BaF_2$ optics, our simulations show that typical manufacturing tolerances on central thickness and radius of curvature only produce a few-percent variation in the maximum kick strength. The situation is expected to further improve in the 0.95-μm configuration due to the use of quartz, which is harder than $BaF_2$ and can be shaped with higher precision. Positioning of the light optics and will be carried out with a commercial hexapod-like motion solution. The chosen system is non-magnetic and meets the vacuum, range and load requirements for all experimental configurations.

**Vacuum chambers:** The vacuum-chamber designs are now being discussed with various manufacturers. In order to reduce magnetic errors in the bypass, we are considering two types of chambers: seam-welded chambers of 316LN steel and extruded or seam-welded aluminium chambers. In the former, the welds may have a slight increase in magnetic permeability [30]. We are performing simulations to examine the magnitude of magnetic errors that can be expected. If the errors are unacceptable, then aluminium chambers with bimetallic flanges can be used instead.

**Other remarks:** We note also that the undulator-radiation measurements will be contaminated to a degree by the synchrotron radiation from the main-ring dipoles. We have simulated this contamination and, at the fundamental frequency, the probability that a detected photon originated from the undulator is ~85%. If an appropriate aperture is applied to the dipole vacuum window then this can be improved to ~95%.

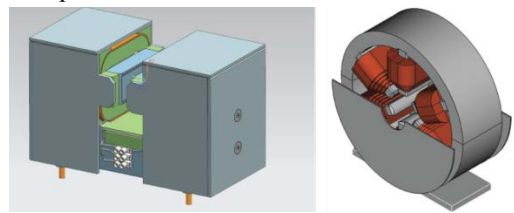

Figure 6: Designs of the OSC chicane dipoles and the IOTA main-ring sextupoles.

## CONCLUSION

We have detailed the conceptual design for the planned demonstration of OSC in Fermilab's IOTA ring, which will constitute the first experimental demonstration of OSC. This program will serve as a pathfinder that explores OSC physics, experimental methods and diagnostics, and it will act as a bridge towards development of the OSC systems required by colliders. At the design wavelength, 0.95 μm, the OSC cooling rate will exceed that from synchrotron radiation damping by a factor of ~60 in the absence of any amplification. The required hardware system, including magnets, light optics and vacuum chambers and beam pipes are maturing in their design and many elements are ready for fabrication or procurement. The installation of the OSC insert in the IOTA ring is planned for spring of 2019 with an expectation of first cooling in summer of the same year.

## ACKNOWLEDGMENTS

Fermilab is managed by the Fermi Research Alliance, LLC for the U.S. Department of Energy Office of Science Contract number DE-AC02-07CH11359.